# Chiral damping of magnons

Dae-Yun Kim[1,2], Imane Berrai[3], T. S. Suraj[4], Yves Roussigné[3], Shuhan Yang[1], Mohamed Belmeguenai[3], Fanrui Hu[1], Guoyi Shi[1], Hui Ru Tan[5], Jifei Huang[4], Anjan Soumyanarayanan[4,5], Kyoung-Whan Kim[6,7], Salim Mourad Cherif[3], and Hyunsoo Yang[1*]

[1]Department of Electrical and Computer Engineering, National University of Singapore, 117583, Singapore

[2]Samsung Advanced Institute of Technology, Samsung Electronics, Suwon, 16678, Republic of Korea

[3]Université Sorbonne Paris Nord, LSPM, CNRS, UPR 3407, F-93430 Villetaneuse, France

[4]Department of Physics, National University of Singapore, Singapore, 117551 Singapore

[5]Institute of Materials Research & Engineering, Agency for Science, Technology & Research (A*STAR), Singapore, 138634 Singapore

[6]Center for Spintronics, Korea Institute of Science and Technology, Seoul, 02792, Republic of Korea

[7]Department of Physics, Yonsei University, Seoul, 03722, Republic of Korea

*eleyang@nus.edu.sg

**Chiral magnets have garnered significant interest due to the emergence of unique phenomena prohibited in inversion-symmetric magnets. While the equilibrium characteristics of chiral magnets have been extensively explored through the Dzyaloshinskii-Moriya interaction (DMI), non-equilibrium properties like magnetic damping have received comparatively less attention. We present the inaugural direct observation of chiral damping through Brillouin light scattering (BLS) spectroscopy. Employing BLS spectrum analysis, we independently deduce both the Dzyaloshinskii-Moriya interaction (DMI) and chiral damping, extracting them from the frequency shift**

**and linewidth of the spectrum peak, respectively. The resulting linewidths exhibit clear odd symmetry with respect to the magnon wave vector, unambiguously confirming the presence of chiral damping. Our study introduces a novel methodology for quantifying chiral damping, with potential ramifications on diverse nonequilibrium phenomena within chiral magnets.**

Inversion symmetry breaking engenders chirality within the system, favoring distinct handedness. This chirality exhibits a widespread presence, having been investigated across diverse academic disciplines, encompassing mathematics [1], physics [2,3], chemistry [4,5], and biology [6]. Among these, the discipline of magnetism stands out as one of the most prominently explored fields concerning chirality, owing to its pivotal significance in contemporary spintronics applications. These applications encompass magnetization switching [7-9], magnetic domain-wall motion [10-12], and the stabilization of magnetic skyrmions [13-16]. In magnetic materials, the breaking of inversion symmetry leads to the emergence of the antisymmetric exchange interaction, commonly referred to as the Dzyaloshinskii-Moriya interaction (DMI) [17,18]. This interaction plays a pivotal role in the generation of magnetic chirality. The DMI not only shapes the spin configuration into a spiral pattern, giving rise to structures like magnetic skyrmions [13,14] or chiral magnetic domain walls (DWs) [19,20], but also imparts a distinct right- or left-handed preference to these spiral spin arrangements. Essentially, the DMI introduces chirality to the magnetic energy landscape. The chirality of magnetic energy is of significant importance, as it not only influences the overall magnetic statics but also holds great promise for applications in the development of next-generation spintronic devices [21-25].

Chirality exerts its influence not only on the equilibrium characteristics of chiral magnets but also extends its impact to nonequilibrium properties, as theoretically postulated [26-29]. Within this context, the chiral component of magnetic dissipation is referred to as chiral damping. In other words, the dependence of the damping of chiral magnets or chiral objects (such as magnetic domain walls, magnetic skyrmions, or magnons) on the handedness of the soliton is called chiral damping. The emergence of chiral damping assumes significance, introducing a novel dimension of magnetic chirality beyond the conventional DMI, and profoundly affecting spin dynamics. In essence, chiral damping is as pivotal as DMI, with the former shaping magnetic chirality within the dynamics of spins, while the latter characterizes it within the static properties. In the context of magnetic DWs and skyrmions, the stability of these magnetic entities is primarily dictated by the DMI [19,20]. Conversely, their dynamic characteristics, including mobility, are influenced by chiral damping. Although numerous efforts have been made to establish the presence of chiral damping, previous findings have primarily relied on indirect approaches. Among these, the measurement of chiral DW speed stands out as a representative method for detecting chiral damping [26,30,31-34]. However, this technique falls short in providing unequivocal evidence of chiral damping due to the following limitations. Firstly, the direct extraction of magnetic damping from domain wall (DW) dynamics poses a significant challenge, making it difficult to achieve a direct observation of chiral damping. Secondly, the asymmetry observed in chiral DW speed, traditionally associated with evidence of chiral damping, may stem from sources beyond chiral damping itself, including factors such as the chiral gyromagnetic ratio [35], chiral DW width [36], or chiral attempt frequency [37]. Lastly, the scope for conducting chiral damping experiments via DW dynamics is limited primarily to specific magnetic films. This limitation arises due to the prerequisite for strong perpendicularly magnetized films with a single-domain phase for effective exploration of chiral DW dynamics.

In this Letter, we present an experimental verification of chiral damping, substantiated through the observation of chirality-dependent magnon dissipation. This phenomenon is assessed via Brillouin light scattering (BLS) spectroscopy, a technique traditionally employed for examining spin-wave dynamics [38,39]. Recently, BLS spectroscopy has garnered considerable interest, particularly for its capacity to quantify the Dzyaloshinskii-Moriya interaction (DMI) [40-43]. Specifically, one can deduce the DMI by comparing the peak frequencies of magnons with opposite handedness. Notably, our work extends the utility of BLS spectroscopy beyond quantifying the chiral energy (DMI) to independently determining chiral damping. Furthermore, the current magnon-based experimental approach offers a direct means of probing magnetic damping, facilitating a clear and unambiguous demonstration of chiral damping. Consequently, our study not only addresses the previously mentioned constraints associated with magnetic domain-wall-based methodologies but also provides valuable support for pioneering research in chiral damping [26-34].

The BLS spectrum encapsulates a wealth of information about magnons. Specifically, the peak frequency within the spectrum mirrors the static attributes of magnons, encompassing their dispersion relation, while the spectrum's linewidth mirrors the dynamic features, notably magnetic damping. In the absence of the DMI, one would typically anticipate the dispersion relation and magnetic damping of magnons to exhibit non-chiral and symmetric behavior with respect to the wave vector ($k$). Contrary to this conventional expectation, our observations reveal asymmetric linewidths in magnons, even in the absence of significant DMI. Furthermore, this linewidth asymmetry demonstrates an anti-symmetric pattern relative to the wave vector, unequivocally indicating the presence of chiral damping. The significance of our findings lies in the direct evidence of chiral damping, which contributes to a more comprehensive understanding of magnetic dynamics. Additionally, the emergence of a new degree of freedom in magnetic chirality holds promise for innovative applications in spintronic devices.

Furthermore, the experimental methodology presented here could serve as a standard technique for the quantification of chiral damping.

Magnetic films exhibiting minimal Dzyaloshinskii-Moriya interaction (DMI) are considered ideal candidates for demonstrating chiral damping. This preference arises from the fact that any observed chiral phenomena in such films can be unequivocally ascribed to the influence of chiral damping, unconfounded by the presence of DMI. For this study, 4 different types of magnetic films are fabricated, each of which structures are MgO (4 nm)/CoFeB (1.2 nm)/MgO (4 nm), Ta (5 nm)/Cu (2.5 nm)/CoFeB (2 nm)/Cu (2.5 nm), Ta (5 nm)/W (2.5 nm)/CoFeB (2 nm)/Cu (2.5 nm), and Ta (5 nm)/Cu (2.5 nm)/CoFeB (2 nm)/W (2.5 nm). To prevent oxidation, a $SiO_2$ (3 nm) layer was capped on these structures. These films are fabricated on $Si/SiO_2$ substrates by using DC/RF magnetron sputtering. For detailed information about sample fabrication, see Section I of Supplemental Material. The first two films lack structural inversion asymmetry (SIA) and thus, negligible DMI is expected in these films. On the other hand, the last two films have clear but opposite SIA with each other.

The BLS spectra are then measured to investigate DMI and magnetic damping of these films. The experimental schematics of BLS spectra measurements is depicted in Fig. 1(a). A strong enough external magnetic field is applied along the $y$-direction so that the magnetization on these films is aligned along the same direction. Note that an external magnetic field of 300 mT was strong enough for the alignment, as all magnetic films exhibit in-plane magnetic anisotropy, as demonstrated by the hysteresis loops (see Section II of Supplemental Material). In this geometry, magnons propagate along the $x$-direction, which is also known as the Damon-Eshbach (DE) mode (please refer to Section of the Supplemental Material) [44]. Figure 1(b) shows the BLS spectrum of DE mode magnons. Stokes and anti-Stokes peaks are clearly observed, each of which corresponds to magnons with the wave vector of $+k$ and $-k$. It should be noticed that these two magnons have opposite handedness as depicted in Fig. 1(c) and (d). In the absence of DMI, symmetry requires these two magnons with opposite handedness to have an identical frequency. On the other hand, in the presence of DMI, the frequencies of these two magnons should be different from each other, since the DMI-induced energy of magnons depends on the handedness of magnons. As shown in Fig. 1(c) and (d), magnons with opposite handedness have opposite sign of DMI-induced energy, $E_{\mathrm{DMI}} = -\vec{D} \cdot (\vec{m}_i \times \vec{m}_j)$, where $D$ is the DMI constant, and $\vec{m}_{i,j}$ is the $i,j$-component of the magnetization. Therefore, in the presence of DMI, the dispersion relation of magnons appears to be asymmetric with respect to $k$.

We first consider the nominally symmetric systems, MgO/CoFeB/MgO and Cu/CoFeB/Cu films in Fig. 2(a) and 2(b), respectively. The dispersion relation of magnons is observed to be symmetric with respect to $k$ as shown in Supplemental Material Section III. Figure 2(c-d) show the difference of magnon frequencies, which is defined as $\Delta f(k) = f(+k) - f(-k)$. In both films $\Delta f$ data points are very scattered, and linear correlations between $\Delta f$ and $k$ are not observed, verifying a negligible DMI (for detailed discussions on non-negligible $\Delta f$, see Section XII of Supplemental Material). On the other hand, the linewidth ($\Gamma$) of magnons is asymmetric with respect to $k$ as shown in Supplemental Material Section III. Figure 2(e-f) plot the linewidth difference of magnons, which is defined as

$$\Delta\Gamma(k) = \Gamma(+k) - \Gamma(-k). \tag{1}$$

In contrast to $\Delta f$, $\Delta\Gamma$ exhibit clear correlation with respect to $k$. In both films, $\Delta\Gamma$ is observed to be directly proportional to $k$. This observation holds significant importance, as the robust linear correlation between $\Delta\Gamma$ and $k$ strongly indicates the presence of magnetic chirality. For the physical variables to have chirality, its spatial distribution (i.e. $k$ dependence) with respect to the symmetric axis should be anti-symmetric (i.e. an odd function of $k$) due to the nature of inversion symmetry breaking. Therefore, the linearity of $\Delta\Gamma(k)$ is the evidence of magnetic chirality. Since these two magnetic films are already demonstrated to have negligible DMI, the observed asymmetric linewidth unambiguously signifies that there exists a new source of magnetic chirality other than DMI.

We examine whether this linewidth asymmetry originates from the chiral damping. The linewidth of Stokes and anti-Stokes peaks corresponds to the imaginary part of the dispersion relation of magnons (i.e. $\Gamma(k) = 2\,\text{Im}(\omega)$, where $\text{Im}(\omega)$ is the imaginary part of the angular frequency). According to Ref. [42, 45],

$$\text{Im}(\omega) = \alpha(k)\gamma\mu_0(H_0 + Jk^2 - H_U + M_S/2)[1 + \omega_{\text{DM}}(k)/\omega_0(k)], \qquad (2)$$

where $\gamma$ is the gyromagnetic ratio, $J$ is the exchange constant, $H_U = 2K/\mu_0 M_S$, $K$ is the magnetic anisotropy energy density, $M_S$ is the saturation magnetization, $\omega_{\text{DM}}(k)$ is the attribution of DMI, and $\alpha(k)$ is the magnetic damping constant. Note that $\omega_{\text{DM}}(k)$ is the odd function of $k$. Referring to Eq. (2), the observed $\Delta\Gamma$ can originate from either DMI-induced term $\omega(k)$ or magnetic damping $\alpha(k)$. For symmetric MgO/CoFeB/MgO and Cu/CoFeB/Cu magnetic films, it is already demonstrated that the DMI is negligible and thus, $\omega(k)$ is also expected to be negligible. Therefore, one can conclude that observed chiral $\Delta\Gamma$ is solely attributed to the magnetic damping $\alpha(k)$, which indicates that magnetic damping of magnons is chiral (see Section XI of Supplemental Material). To the best of our knowledge, this is the first direct observation of chiral damping.

The SIA serves as the fundamental origin of magnetic chirality (both DMI and chiral damping). However, the microscopic mechanisms by which DMI and chiral damping arise from SIA differ. In the case of DMI, the impact of SIA on the Fermi-sea of local magnetic moments determines the strength of DMI. In the case of chiral damping, the influence of SIA on the Fermi-surface of conduction electrons (or any entities interacting with local magnetic moments) determines the strength of chiral damping. Therefore, even though chiral damping may be significant, DMI can be negligible, and vice versa.

For further studies on chiral damping, we investigate the magnons frequency and linewidth for asymmetric magnetic films, as shown in Fig. 3. Similar to symmetric MgO/CoFeB/MgO and Cu/CoFeB/Cu films, asymmetric W/CoFeB/Cu and Cu/CoFeB/W films in Fig. 3(a) and 3(b), respectively, are determined to have negligible DMI because the difference of magnon frequency, $\Delta f$, does not exhibit a linear correlation with $k$ as shown in Fig. 3(c) and (d). On the other hand, a clear linear correlation between $\Delta\Gamma$ and $k$ is observed in the W/CoFeB/Cu film as shown in Fig. 3(e). Therefore, one can conclude that the chiral damping is also manifested in the W/CoFeB/Cu film.

In contrast to the other magnetic films, a clear but non-linear correlation between $\Delta\Gamma$ and $k$ is observed in the Cu/CoFeB/W film as shown in Fig. 3(f). This non-linear correlation may be attributed to a higher order contribution of chiral damping. As aforementioned, for the source of $\Delta\Gamma$ to be the magnetic chirality, $\Delta\Gamma$ should be an odd-function of the $k$ vector ($\sum_{n=0}^{\infty} \eta_{2n+1} k^{2n+1}$). As the simplest and the most probable case, we check whether the first two lowest-order term ($\eta_1 k + \eta_3 k^3$) accounts for the non-linear correlation between $\Delta\Gamma$ and $k$. The solid curve in Fig. 3(f) presents the best fitting of $\eta_1 k + \eta_3 k^3$. Therefore, the observed $\Delta\Gamma(k)$ of Cu/CoFeB/W film is attributed to the magnetic chirality. Since this film is already determined to have negligible DMI, one can conclude that the non-linear $\Delta\Gamma(k)$ of Cu/CoFeB/W film is the consequences of higher-order contribution of chiral damping. This can originate from higher-order non-local contributions of the degrees of freedom mediating the chiral damping, such as electrons and phonons. Previous theories of chiral damping account for the lowest-order non-local contribution $\sim \nabla \mathbf{m}$, capturing terms of the first order in $k$. However, our result shows the importance of considering higher-order non-local contributions, which has been largely overlooked so far.

To provide the most straightforward and unambiguous demonstration of chiral damping, our investigation focuses on magnetic films characterized by negligible chiral energy (DMI). It is noteworthy that in more general scenarios, where both chiral energy (DMI) and chiral damping coexist, one can still infer chiral damping from the linewidth of BLS spectra. For details, please refer to Supplemental Material Section V.

As the next step, we compare the magnitude of chiral damping across four different films. To facilitate this comparison, we extract a non-chiral damping term ($\alpha_0$) and a chiral damping term ($\alpha_C$) following ref. [26], which are defined as the symmetric (even) and antisymmetric (odd) components of $\Gamma(k)$, respectively. Specifically, we extract $\alpha_0$ and $\alpha_C$ from the $0^{th}$ and $1^{st}$ order term of $k$ in $\Gamma(k)$, respectively. Figure 4 shows the ratio $\alpha_C/\alpha_0$ for a given value of $k = 20$ rad/μm for the four different films. The ratio $\alpha_C/\alpha_0$ for the last two asymmetric films is larger than those for the first two symmetric films, supporting a strong correlation between structural inversion asymmetry and chiral damping. The observation of chiral damping in MgO/CoFeB/MgO and Cu/CoFeB/Cu films can be attributed to their microscopic inversion-asymmetric characteristics despite their nominally inversion-symmetric stack structures [48-50]. As shown in Figure S4 of the Supplemental Material, the energy dispersive X-ray spectroscopy demonstrates that the spatial distribution of elemental intensities exhibits inversion asymmetry in all films.

In summary, we have observed asymmetrical magnon linewidths, a phenomenon attributed to magnetic chirality. Our conclusive findings demonstrate that this linewidth asymmetry stems from chiral energy dissipation (chiral damping), rather than the chiral energy (DMI). To the best of our knowledge, this constitutes the first direct experimental confirmation of chiral damping. Our present study bears significant importance, as magnetic damping exerts a substantial influence on the overarching spin dynamics, akin to how DMI shapes the broader landscape of spin statics. Additionally, we foresee that our approach for measuring chiral damping lays a robust foundation for future investigations employing Brillouin light scattering (BLS) spectroscopy to study chiral damping without involving complex devices. From an application point of view, magnetic damping significantly impacts the operational speed and power consumption of devices. Therefore, our direct verification of chirality as a new degree of freedom in magnetic damping opens possibilities for more effective device engineering.

This research was supported by UParis-NUS 2023 award (Inducing magnetic chirality without heavy metals), Merlion Programme 2024 (Spintronics of Low Symmetry Magnetic Layers). K.-W. K. acknowledges financial support from the National Research Foundation of Korea (RS-2024-00334933) and the KIST Institutional Program. This work was supported by Singapore's RIE2025 Manufacturing, Trade and Connectivity (MTC) initiative (A*STAR IRG Grant No. M23M6c0112), and the Ministry of Education Academic Research Fund (AcRF Tier-1 NUS Grant No. 23-1072-A0001).

**References**

[1] M. Petitjean, Optim. Lett. **14**, 329 (2020)

[2] S. Zhang, Y.-S. Park, J. Li, X. Lu, W. Zhang, and X. Zhang, Phys. Rev. Lett. **102**, 023901 (2009)

[3] V. A. Fedotov, P. P. Mladyonov, S. L. Prosvirnin, A. V. Rogacheva, Y. Chen, and N. I. Zheludev, Phys. Rev. Lett. **97**, 167401 (2006)

[4] W. Ma, L. Xu, A. F. de Moura, X. Wu, H. Kuang, C. Xu, and N. A. Kotov, Chem. Rev. **117**, 8041 (2017)

[5] W. Xiao, K.-H. Ernst, K. Palotas, Y. Zhang, E. Bruyer, L. Peng, T. Greber, W. A. Hofer, L. T. Scott, and R. Fasel, Nat. Chem. **8**, 326 (2016)

[6] T. Suzuki, Y. Kosugi, M. Hosaka, T. Nishimura, and D. Nakae, Environmental Toxicology and Chemistry **33**, 2671 (2014)

[7] I. M. Miron, K. Garello, G. Gaudin, P.-J. Zermatten, M. V. Costache, S. Auffret, S. Bandiera, B. Rodmacq, A. Schuhl, and P. Gambardella, Nature **476**, 189 (2011)

[8] L. Liu, O. J. Lee, T. J. Gudmundsen, D. C. Ralph, and R. A. Buhrman, Phys. Rev. Lett. **109**, 096602 (2012)

[9] L .Liu, C.-F. Pai, Y. Li, H. W. Tseng, D. C. Ralph, and R. A. Buhrman, Science **336**, 6081 (2012)

[10] I. M .Miron, T. Moore, H. Szambolics, L. D. Buda-Prejbeanu, S. Auffret, B. Rodmacq, S. Pizzini, J. Vogel, M. Bonfim, A. Schuhl, and G. Gaudin, Nat. Mater. **10**, 419 (2011)

[11] P. P. J. Haazen, E. Jue, J. H. Franken, R. LAvrijsen, H. J. M. Swagten, and B. Koopmans, Nat. Mater. **12**, 299 (2013)

[12] K.-S. Ryu, L. Thomas, S.-H. Yang, and S. Parkin, Nat. Nanotechnol. **8**, 527 (2013)

[13] U. K. Robler, A. N. Bogdanov, C. Pfleiderer, A. Rosch, A. Neubauer, R. Georgii, and P. Boni, Science **323**, 915 (2009)

[14] X. Z. Yu, Y. Onose, N. Kanazawa, J. H. Park, J. H. Han, Y. Matsui, N. Nagaosa, and Y. Tokura, Nature **465**, 901 (2010)

[15] N. Nagaosa and Y. Tokura, Nat. Nanotechnol. **8**, 899 (2013)

[16] A. Fert, V. Cros, and J. Sampaio, Nat. Nanotechnol. **8**, 152 (2013)

[17] T. Moriya, Phys. Rev. **120**, 91 (1960)

[18] I. Dzyaloshinskii, J. Phys. Chem. Solids **4**, 241 (1958)

[19] M. Heide, G. Bihlmayer, and S. Blugel, Phys. Rev. B **78**, 140403(R) (2008)

[20] A. Thiaville, S. Rohart, E. Jue, V. Cros, and A. Fert, Europhys. Lett. **100**, 57002 (2012)

[21] S. S. P. Parkin, M. Hayashi, and L. Thomas, Science **320**, 5873 (2008)

[22] K. L. Wang, J. G. Alzate, and P. K. Amiri, J. Phys. D: Appl. Phys. **46**, 74003 (2013)

[23] I. Ahmed, Z. zhao, M .G. Mankalale, S. S. Sapatenekar, J.-P. Wang, and C. H. Kim, IEEE J. Explor. Solid-State Comput. Devices Circuits **3**, 74 (2017)

[24] H.-S. P. Wong and S. Salahuddin, Nat. Naotechnol. **10**, 191 (2015)

[25] S. Jung, H. Lee, S. Myung, H. Kim, S. K. Yoon, S.-W. Kwon, Y. Ju, W. Yi, B. Kwon, B. Seo, K. Lee, G.-H. Koh, K. Lee, Y. Song, C. Choi, D. Ham, and S. J. Kim, Nature **601**, 211 (2022)

[26] E. Jue, C. K. Safeer, M. Drouard, A. Lopez, P. Balint, L. Buda-Prejbeanu, O. Boulle, S. Auffret, A. Schuhl, A. Manchon, I. M. Miron, and G. Gaudin, Nat. Mater. **15**, 272 (2016)

[27] C. A. Akosa, I. M. Miron, G. Gaudin, and A. Manchon, Phys. Rev. B **93**, 214429 (2016)

[28] K.-W. Kim and H.-W. Lee, Nat. Mater. **15**, 25 (2016)

[29] F. Freimuth, S. Blugel, and Y. Mokrousov, Phys. Rev. B **96**, 104418 (2017)

[30] C. K. Safeer, M.-A. Nsibi, J. Nath, M. S. Gabor, H. Yang, I. Joumard, S. Auffret, G. Gaudin, and I.-M. Miron, Nat. Commun. **13**, 1192 (2022)

[31] A. Ganguly, S. Zhang, I. M. Miron, J. Kosel, X. Zhang, A. Manchon, N. Singh, D. H. Anjum, and G. Das, Appl. Electron. Mater **3**, 4734 (2021)

[32] A. V. Davydenko, A. G. Kozlov, M. E. Stebliy, A. G. Kolesnikov, N. I. Sarnavskiy, I. G. Iluishin, and A. P. Golikov, Phys. Rev. B **103**, 094435 (2021)

[33] R. Lavrijsen, D. M. F. Hartmannn, A. van den Brink, Y. Yin, B. Barcones, R. A. Duine, M. A. Verheijen, H. J. M. Swagten, and B. Koopmans, Phys. Rev. B **91**, 104414 (2015)

[34] D. -Y. Kim, M.-H. Park, Y.-K. Park, J.-S. Kim, Y.-S. Nam, D.-H. Kim, S.-G. Je, H.-C. Choi, B.-C. Min, and S.-B. Choe, NPG Asia Mater. **10**, e464 (2018)

[35] K.-W. Kim, H.-W. Lee, K.-J. Lee, K. Everschor-Sitte, O. Gomonay, and J. Sinova, Phys. Rev. B **97**, 100402(R) (2018).

[36] D.-Y. Kim, D.-H. Kim, and S.-B. Choe, Appl. Phys. Express **9**, 053001 (2016)

[37] S. Lemerle, J. Ferre, C. Chappert, V. Mathet, T. Giamarchi, and P. Le Doussal, Phys. Rev. Lett. **80**, 849 (1998)

[38] B. Hillebrands, Phys. Rev. B **41**, 530 (1990).

[39] J. Jorzick, S. O. Demokritov, C. Mathieu, B. Hillebrands, B. Bartenlian, C. Chappert, F. Rousseaux, and A. N. Slavin, Phys. Rev. B **60**, 15194 (1999)

[40] H. T. Nembach, J. M. Shaw, M. Weiler, E. Jué & T. J. Silva, Nat. Phys. **11**, 825 (2015).

[41] J. Cho, N-H. Kim, S. Lee, J-S. Kim, R. Lavrijsen, A. Solignac, Y. Yin, D-S. Han, N. J. J. van Hoof, H. J. M. Swagten, B. Koopmans and C-Y. You, Nat. Commun. **6**, 7635 (2015)

[42] K. Di, V. L. Zhang, H. S. Lim, S. C. Ng, M. H. Kuok, J. Yu, J. Yoon, X. Qiu, and H. Yang, Phys. Rev. Lett. **114**, 047201 (2015)

[43] M. Belmeguenai, J.-P. Adam, Y. Roussigné, S. Eimer, T. Devolder, J-V. Kim, S.M. Chérif, A.A. Stashkevich, A. Thiaville, Phys. Rev. B **91**, 180405(R) (2015).

[44] R. W. Damon and J. R. Eshbach, J. Appl. Phys. **31**, S104 (1961).

[45] J.-H. Moon, S.-M. Seo, K.-J. Lee, K.-W. Kim, J. Ryu, H.-W. Lee, R. D. McMichael, and M. D. Stiles, Phys. Rev. B **88**, 184404 (2013)

[46] G. Feng, S. van Dijken, J. F. Feng, J. M. D. Coey, T. Leo, and D. J. Smith, J. Appl. Phys. **105**, 033916 (2009)

[47] J. Y. Bae, W. C. Lim, H. J. Kim, T. D. Lee, K. W. Kim, and T. W. Kim, J. Appl. Phys. **99**, 08T316 (2006)

[48] R. M. Rowan-Robinson, A. A. Stashkevich, Y. Roussigné, M. Belmeguenai, S.-M. Chérif, A. Thiaville, T. P. A. Hase, A. T. Hindmarch, and D. Atkinson, Sci. Rep. **7**, 16835 (2017)

[49] M. Belmeguenai, J.-P. Adam, Y. Roussigné, S. Eimer, T. Devolder, J.-V. Kim, S. M. Cherif, A. Stashkevich, and A. Thiaville, Phys. Rev. B **91**, 180405(R) (2015)

[50] O. Inyang et al., Appl. Phys. Lett. **123**, 122403 (2023)

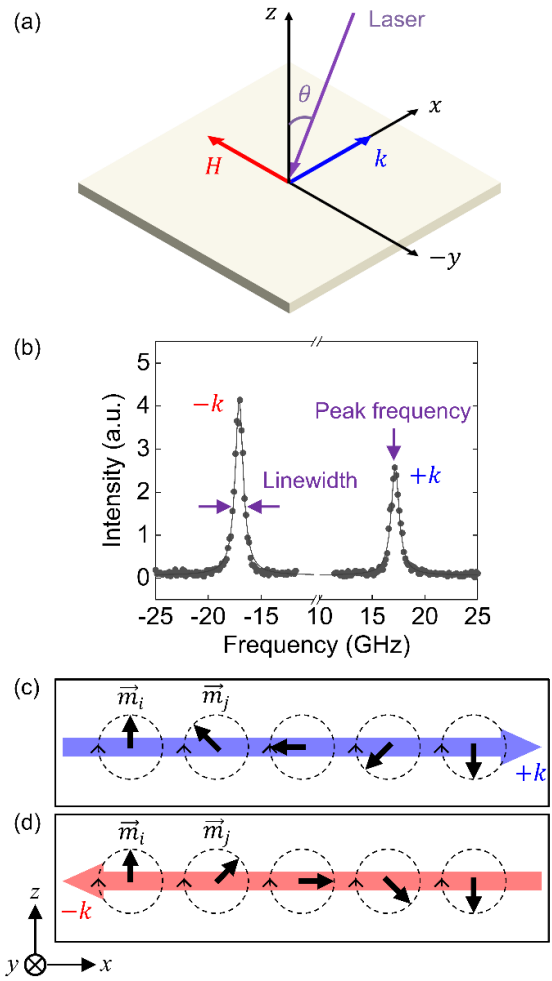

Figure 1. (a) Schematic diagram of BLS measurements. External magnetic field $H$ is applied along the *y*-direction and the laser is incident along the *xz*-plane. (b) BLS spectrum of Cu/CoFeB/Cu film for $k = 11.8$ μm$^{-1}$. (c,d) Propagating magnons with wave vectors of $+k$ (c) and $-k$ (d). Dashed circle presents precession of local magnetization, $\vec{m}_{i,j}$ (bold arrow). Colored arrows describe propagation directions of magnons.

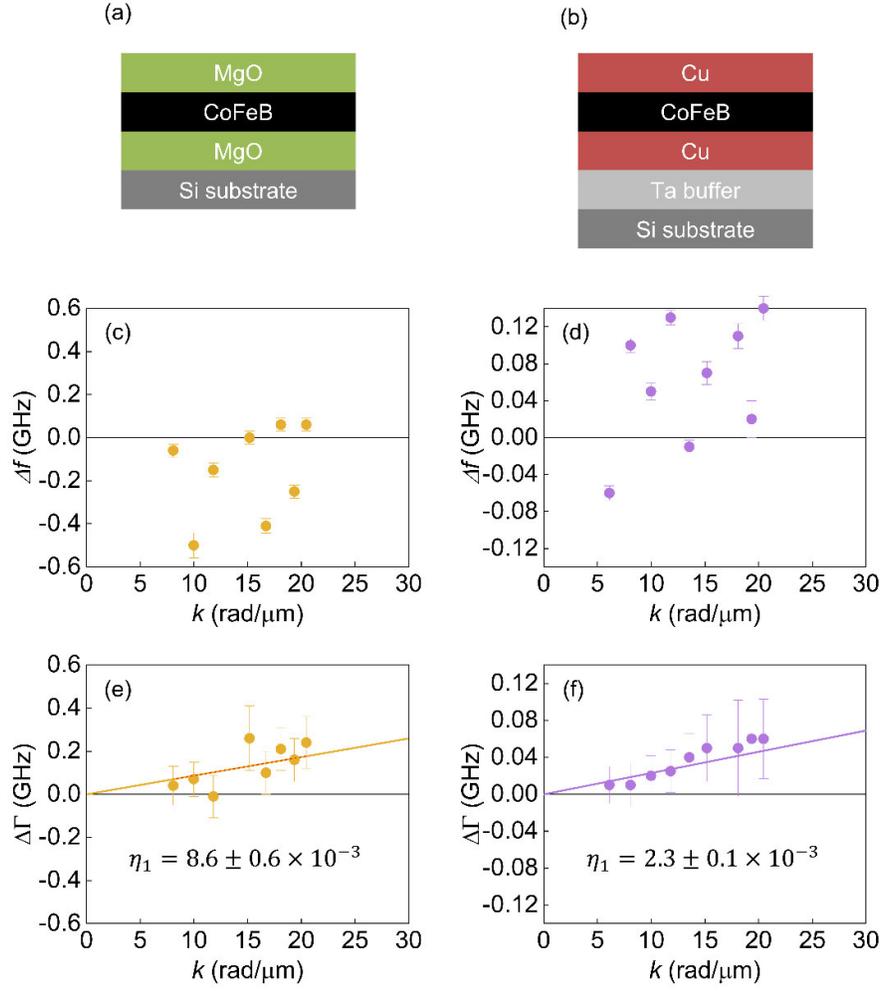

Figure 2. (a,b) Layered structures of symmetric MgO/CoFeB/MgO (a) and Cu/CoFeB/Cu (b) films. (c,d) Plots of magnon frequency difference ($\Delta f$) versus wave vector ($k$) for MgO/CoFeB/MgO (c) and Cu/CoFeB/Cu (d) films. (e,f) Plots of $\Delta\Gamma$ versus $k$ for MgO/CoFeB/MgO (e) and Cu/CoFeB/Cu (f) films. Solid lines represent linear fits passing through the origin (see Section VII of Supplemental Material for details).

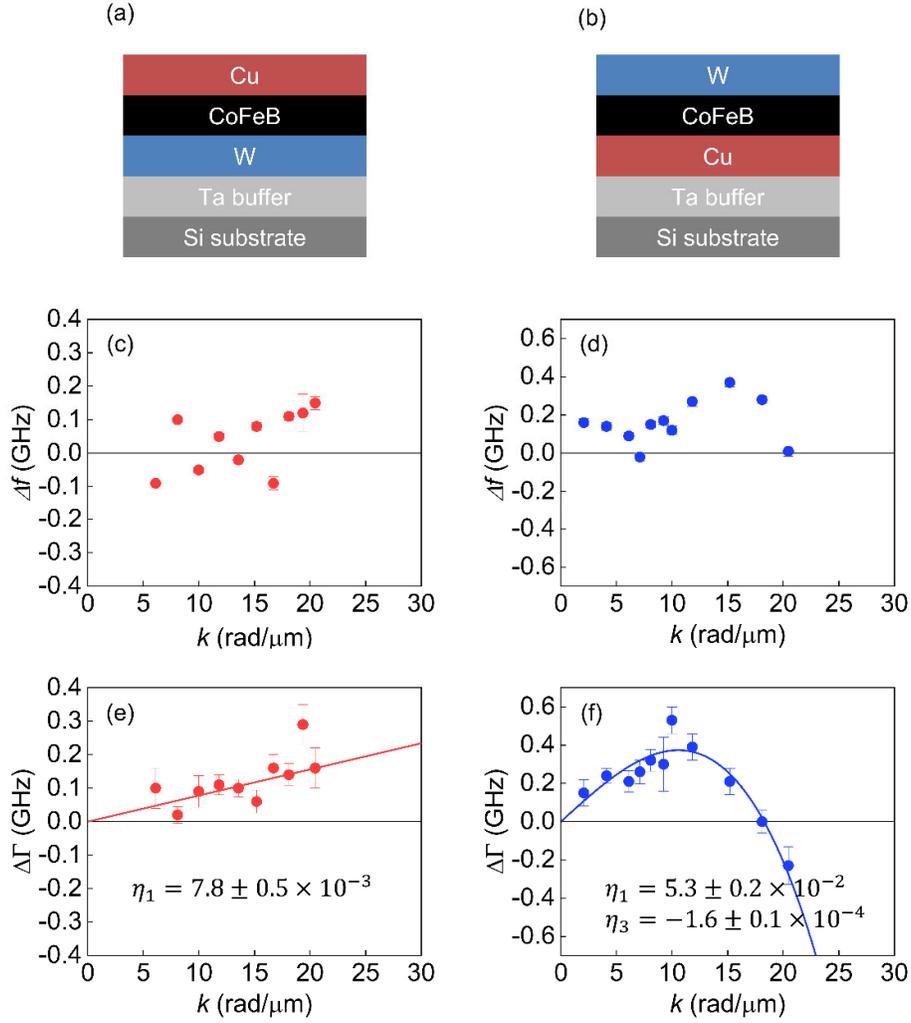

Figure 3. (a,b) Layered structures of asymmetric W/CoFeB/Cu (a) and Cu/CoFeB/W (b) films. (c,d) Plots of magnon frequency difference ($\Delta f$) versus wave vector ($k$) for W/CoFeB/Cu (c) and Cu/CoFeB/W (d) films. (e,f) Plots of $\Delta\Gamma$ versus $k$ for W/CoFeB/Cu (e) and Cu/CoFeB/W (f) films. Red solid line presents the best linear fit. Blue solid curve is the best fit using $\Delta\Gamma = \eta_1 k + \eta_3 k^3$.

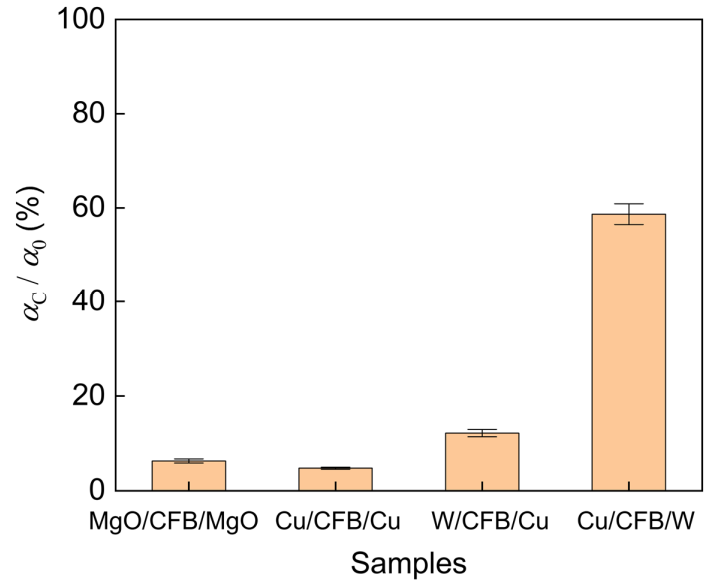

Figure 4. Plot of $\alpha_C/\alpha_0$ for $k = 20$ rad/μm.